# Analysis of the residual force noise for the LISA Technology Package

**Luigi Ferraioli[1,7], Michele Armano[2], Giuseppe Congedo[1], Marc Diaz-Aguilo[3], Fabrizio De Marchi[1], Adrien Grynagier[4], Martin Hewitson[5], Mauro Hueller[1], Anneke Monsky[5], Miquel Nofrarias[5], Eric Plagnol[6], Boutheina Rais[6] and Stefano Vitale[1].**

[1]University of Trento and INFN, via Sommarive 14, 38123 Povo (Trento), Italy

[2]SRE-SD ESAC, European Space Agency, Camino bajo del Castillo s/n, Urbanizacin Villafranca del Castillo, Villanueva de la Cañada, 28692 Madrid, Spain

[3]UPC/IEEC, EPSC, Esteve Terrades 5, E-08860 Castelldefels, Barcelona, Spain

[4]Institut für Flugmechanik und Flugregelung, 70569 Stuttgart, Germany

[5]Albert-Einstein-Institut, Max-Planck-Institut für Gravitationsphysik und Universität Hannover, 30167 Hannover, Germany

[6]APC UMR7164, Université Paris Diderot, Paris, France.

E-mail: luigi@science.unitn.it

**Abstract**. The analysis of the noise sources perturbing a test mass (TM) geodesic motion is the main scientific objective of the LISA Technology Package experiment (LTP) on board of the LISA Pathfinder space mission. Information on force noise acting on TMs are obtained with a data reduction procedure involving system parameters. Such parameters can be estimated from dedicated experimental runs. Therefore the final estimation of force noise is affected by two sources of uncertainty. One is statistical and connected to the random nature of noisy signals. The other is connected to the uncertainties on the system parameters. The analysis of simulated LTP data is indicating that the major contribution to the force noise power spectral density uncertainties is coming from the statistical properties of the spectrum estimator.

## 1. Introduction

The LISA Technology Package (LTP) experiment is the scientific payload of the European Space Agency mission, LISA Pathfinder. Its goal is to determine and analyse all possible sources of disturbance which perturb a free-falling test masses (TMs) from its geodesic motion [1 - 4]. The system is composed of two test masses whose position is sensed by an interferometer. The spacecraft cannot simultaneously follow both masses, therefore the trajectory of only one test mass serves as the drag-free reference along the *x* (measurement) axis. To prevent the trajectories of the two masses from diverging in response to any differential force, the second test mass is electrically actuated to follow the first test mass. In the main science operating mode, the position of the Spacecraft (SC) relative to the first test mass is controlled using micro-Newton thrusters attached to the SC. The position of the second test mass is controlled using capacitive actuators surrounding the test masses. The first interferometer channel measures the position of the spacecraft relative to the first test mass. The second interferometer channel (differential channel) measures the relative displacement between the two test masses.

---
[7] To whom any correspondence should be addressed.

One of the primary mission outcomes will be the estimation of the residual force per unit of mass perturbing the motion of the free falling test mass. Since the instrument provides displacement data, the effective force per unit of mass is obtained by an offline reduction procedure. Such a procedure is based on a model of the system which is characterized by some dynamics parameters such as the test mass stiffness (coupling of TMs to the spacecraft), the gain of the control loops, the interferometer cross-talks and various bus delays (see table 1). The values of the parameters and their uncertainties will be determined with dedicated on-flight experiments, therefore we expect the final estimation of the residual force per unit of mass being affected by parameters uncertainties. The work reported in the present paper demonstrates that the uncertainty on the estimated noise force introduced by parameter estimation is negligible with respect to the uncertainty connected with the statistical properties of the power spectral density estimators. We performed our calculation with the data analysis tools provided by the LTPDA toolbox [5].

## 2. LTP, the LISA Technology Demonstrator

The dynamical equations assumed for test mass 1 (TM$_1$) and test mass 2 (TM$_2$), along the $x$ axis of the interferometer can be written in Laplace notation as:

$$\begin{aligned}
\text{TM}_1: \quad & s^2 x_1 + \omega_1^2 \left(1 + \frac{m_1}{m_{sc}}\right) x_1 + \frac{m_2}{m_{sc}} \omega_2^2 x_1 + \frac{m_2}{m_{sc}} \omega_2^2 x_\Delta + \Gamma_{21} x_\Delta = \cdots \\
& \cdots \frac{F_{1e}}{m_1} + \left(\frac{m_{sc} + m_1}{m_{sc} m_1}\right) F_{1i} + \frac{F_{2i}}{m_2} - \frac{F_{sc}}{m_{sc}} + \frac{F_{21}}{m_1} + \frac{1}{m_{sc}} H_{sus}(s) o_\Delta - \frac{1}{m_{sc}} H_{df}(s) o_1 \\
\text{TM}_2: \quad & s^2 x_\Delta + \omega_2^2 x_\Delta + \left(\omega_2^2 - \omega_1^2\right) x_1 - \left(\Gamma_{12} + \Gamma_{21}\right) x_\Delta = \cdots \\
& \cdots \frac{F_{2e} + F_{2i}}{m_2} - \frac{F_{1e} + F_{1i}}{m_1} - \left(\frac{m_1 + m_2}{m_1 m_2}\right) F_{21} + \frac{1}{m_2} H_{sus}(s) o_\Delta.
\end{aligned} \quad (1)$$

TM$_1$ is assumed to be free falling. In equation (1),

- $\omega_1^2$ and $\omega_2^2$ are the effective coupling of TM$_1$ and TM$_2$ to the Spacecraft.
- $x_1$ is the SC position with respect to the TM$_1$ and $x_\Delta$ is the relative position of the two test masses.
- $F_{1e}$ are external forces acting on TM$_1$, $F_{1i}$ are forces acting on TM$_1$ from the inside of the spacecraft. Analogously $F_{2e}$ and $F_{2i}$ are external and internal forces acting on TM$_2$. $F_{sc}$ are external force acting on the spacecraft, whereas F$_{21}$ are the forces mutually exchanged by the TMs.
- $m_1$, $m_2$ and $m_{sc}$ are TM$_1$, TM$_2$ and spacecraft masses respectively.
- $\Gamma_{21}$ is the first order term of the Taylor series expansion of the gravity force exerted by TM$_2$ on TM$_1$. In analogy $\Gamma_{12}$ is the first order term of the gravity force acted by TM$_1$ on TM$_2$. Since the two TMs masses are approximately equal then we have $\Gamma_{21} \approx \Gamma_{12}$. The order zero of the gravity forces between TMs are constant terms and they are included in $F_{21}$.
- $H_{df}(s)$ and $H_{sus}(s)$ are the drag-free and electrostatic-suspension control laws respectively [3, 4].

- $o_1$ and $o_\Delta$ are the interferometer readout channels sensing $x_1$ and $x_\Delta$ respectively.

## 3. Estimating Force Noise on Test Masses

The main contributions to the effective acceleration noise power spectral density (PSD) on the two interferometer channels can be calculated on the basis of the different noise sources expected at the input:

$$S_{a_1} \approx \frac{1}{m_{sc}^2} S_{sc} + \frac{1}{m_1^2} S_1 + \left(s^2 + \omega_1^2\right)^2 S_{orn_1}$$

$$S_{a_\Delta} \approx \frac{1}{m_1^2} S_1 + \frac{1}{m_2^2} S_2 + \left(s^2 + \omega_2^2\right)^2 S_{orn_\Delta} + \left(\omega_2^2 - \omega_1^2\right)^2 S_{orn_1} \quad (2)$$

Where $S_{sc}$ is the spacecraft force noise PSD, $S_1$ is the TM$_1$ force noise PSD, $S_{orn1}$ is the interferometer readout noise PSD on channel 1, $S_2$ is the TM$_2$ force noise PSD and $S_{orn12}$ is the interferometer readout noise PSD on the differential channel. The output of the first interferometer channel (first line in equation (2)) is dominated by the force noise on the spacecraft. Such noise is mainly due to thrusters noise, solar wind and asymmetric black-body emission from the spacecraft faces. Interferometer readout noise is shaped by the in-loop dynamics which is adding a $f^2$ trend ($f$ is the frequency). Therefore it is of importance only in the high frequency region while the contribution in the frequency range of interest for LISA Pathfinder (~10$^{-3}$ Hz) is negligible. At the output of the differential channel (second line in equation (2)) dominates the contribution from the combined force noise on the two test masses. As already stated, interferometer readout noise contribution is effective only at high frequencies. A projection of the different contribution on the output of the differential channel is reported in figure 1. We used the input noise shapes defined in the requirements for the noise budget of the LISA Pathfinder mission. They effectively correspond to a worst case scenario, therefore we expect better performances during the mission operations. From the analysis of equation (2) and figure 1 we can conclude that the best estimation of TMs force noise is provided by the differential channel output. Force contributions on the two TMs cannot be disentangled.

## 4. System Parameters Estimation

System parameters are estimated with a fit procedure on a linearized model of LTP. The linearization is performed in terms of the physical parameters. Many parameters are barely physically distinguishable each other, therefore fit matrix is often rank deficient. The problem is overcome with singular value decomposition (svd) which identifies a new set of linearly independent parameters. Such new parameters are indeed linear combinations of the physical parameters. The fit is performed for the linearly independent parameters on each available data series. If some physical parameters are indistinguishable then the number of fit parameters is lower than the number of physical parameters, therefore a single measurements is not sufficient to extract all the physical information. We developed a procedure for the combination of different experiments. For each experiment we perform a fit on the svd parameters, then, at the end of the process, we solve the system of linear equations connecting physical and svd parameters, the values for the svd parameters are weighted with the errors obtained by the fits. In this way the information from different experiments is joined to get the final estimation of the full set of physical parameter. In order to ensure the procedure is successful the different experiments must be carefully designed. Each experiment must provide information on a subset of the physical parameters, the union of the subsets must contain the complete set of the physical parameters. In the example reported below, we performed two experiments. The first experiment is injecting a train of sine waves at different frequencies in the guidance of the spacecraft drag-free controller. The output is taken on the two $x$ channels of the interferometer. The signal on the $o_1$ channel provides information on $G_{df}$, $\omega_1^2$, $\tau_{th}$ and $\Delta T_1$. The signal on the differential channel allows the accurately determination of the difference of stiffness on TMs $\omega_2^2 - \omega_1^2$ and the interferometer cross-talk $S_{21}$. The second experiment is injecting a train of sine waves of different frequency on the guidance input of the electric suspension controller. In this experiment we obtain no signal at the output $o_1$. This is indicating that the differential motion of the two TMs is not leaking in the interferometer first channel. The signal on $o_\Delta$ provides information on the parameters $G_{sus}$, $\omega_2^2$, $\tau_{sus}$ and $\Delta T_\Delta$. Further details on the experiments can be found in [6, 7]. The estimated physical parameters values and their descriptions are reported in table 1.

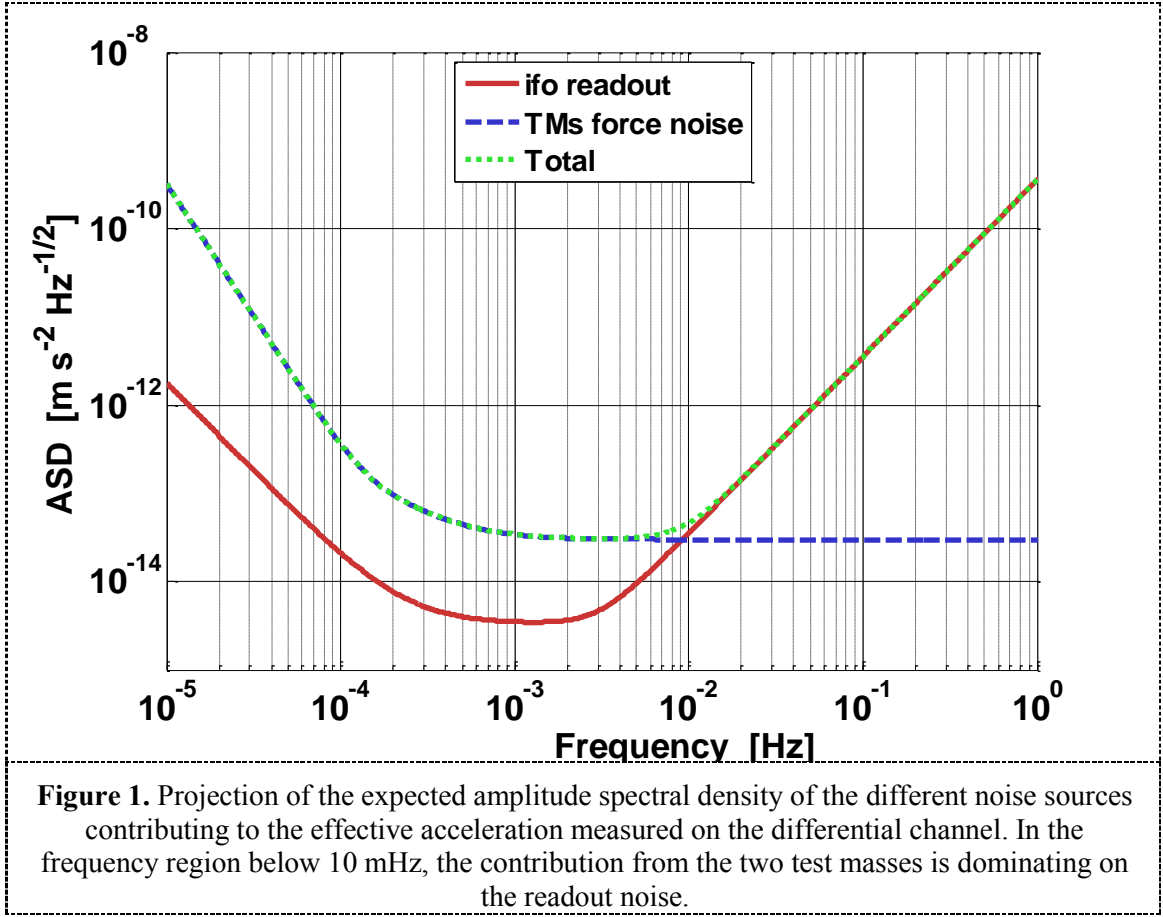

**Figure 1.** Projection of the expected amplitude spectral density of the different noise sources contributing to the effective acceleration measured on the differential channel. In the frequency region below 10 mHz, the contribution from the two test masses is dominating on the readout noise.

**Table 1:** Parameters values obtained by the fit procedure

| Parameter | Value | Description |
|---|---|---|
| $G_{df}$ | $1.0813 \pm 0.0005$ | Drag Free controller gain |
| $G_{sus}$ | $1.0000 \pm 0.0001$ | Suspension controller gain |
| $S_{21}$ | $(1.15 \pm 0.4) \times 10^{-6}$ | Interferometer cross-talk $x_1 \rightarrow o_\Delta$ |
| $\omega_1^2$ | $(-1.319 \pm 0.004) \times 10^{-6}$ [s$^{-2}$] | TM1 global stiffness along x |
| $\omega_2^2$ | $(-2.035 \pm 0.004) \times 10^{-6}$ [s$^{-2}$] | TM2 global stiffness along x |
| $\tau_{th}$ | $0.417 \pm 0.002$ [s] | Characteristic time of thrusters actuation |
| $\tau_{sus}$ | $0.201 \pm 0.003$ [s] | Characteristic time of suspension actuation |
| $\Delta T_1$ | $0.1997 \pm 0.0003$ [s] | Delay on $o_1$ |
| $\Delta T_\Delta$ | $0.1997 \pm 0.0003$ [s] | Delay on $o_\Delta$ |

## 5. Confidence Intervals for Force Noise Estimation

Noise data series are produced with the mission simulator which was designed by industry for testing and validating the drag-free control system and has now been extended to act as a performance simulator that we can use to validate the planned experiments at the tele-command level. We developed a procedure for the extraction of the effective forces per unit of mass from the outputs of the interferometer along the $x$ coordinate ($o_1$ and $o_\Delta$). Since the LTP is a controlled system, we calculate the force commanded by the controllers on the TMs on the basis of the $o_1$ and $o_\Delta$ signals. We used infinite impulse response filter models for the controllers. Then we convert $o_1$ and $o_\Delta$ in physical coordinates $x_1$ and $x_\Delta$ and we apply the dynamics reported in equation (1). The calculation is performed in time domain with the assumption that $s^2 \to d^2/dt^2$. The second derivative is approximated with a five points formula as reported in detail in [8]. Such a procedure allows estimating the effective force per unit of mass acting on the TMs, the estimated force is only corrupted by the interferometer readout noise as shown in equation (2). Further details on the conversion of LTP displacement data to effective force per unit of mass can be found in [7]. The estimation of the effective acceleration is affected by two sources of uncertainties. One is statistical and caused by the random nature of the noisy signal, the other is connected with the fit procedure. Fit results are, indeed, used into the dynamical model for the conversion of displacement data to force per unit of mass, thus the uncertainties on the fit parameters is affecting the estimation of the effective force acting on the test masses.

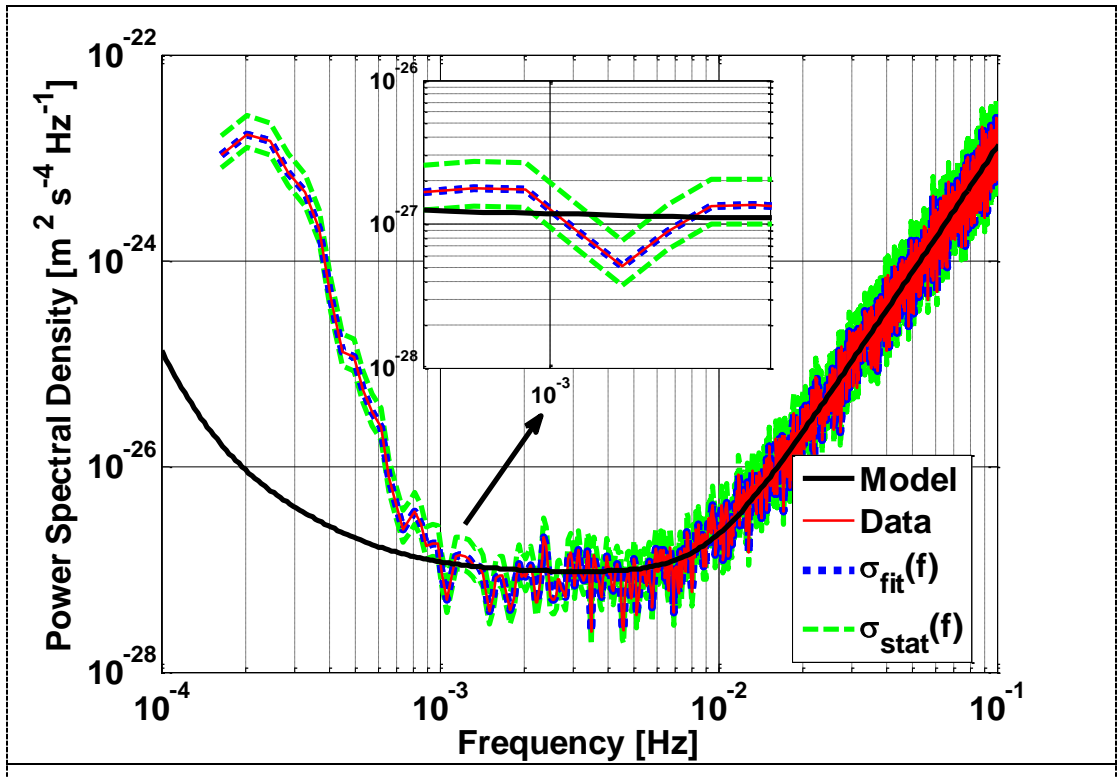

**Figure 2.** Power spectral density of effective acceleration for the interferometer differential channel (red trace) and theoretical model (black curve). Blue dot traces are 68% confidence levels propagated from fit parameters uncertainty. Green dash traces are confidence levels 68% from the statistics of the spectrum estimator. The inset is a zoom on the frequency region around 1 mHz.

Assuming that the parameters are normally distributed around the best fit value, we run a Montecarlo simulation for the estimation of the confidence level on the force estimation. We run $N =$

1000 conversions to force per unit of mass, each time the values of the parameters are randomly generated according to their statistics. It is worth noting that the displacement data are always the same, only the parameters of the dynamics are changing at each iteration. We calculate the spectrum of each of the N force data series and then the sample variance for each frequency point σ$_{fit}$(f) that is assumed defining a 68% confidence level on the estimated force noise spectrum arising from the uncertainty on dynamics parameters. Together with σ$_{fit}$(f) there is another source of uncertainty on the estimation of the force noise spectrum, it is connected with the statistical properties of the spectral estimator[8] and we indicate it with σ$_{stat}$(f). We estimated the spectra with the windowed averaged periodogram method, with 8 averages and 50% overlap between different segments. Each segment is linearly de-trended in time domain and multiplied by a 4-term Blackman-Harris window. The 68% confidence interval for our spectral estimation is then calculated with the procedure described in [11]. The results are reported in figure 2, where we show two different confidence intervals. The blue-dot trace is the confidence interval originated by the propagation of the fit parameters uncertainty. Such errors are so close to the data (red trace) that cannot be distinguished by them. The green-dash trace is the confidence interval defined by the statistical properties of the power spectral density estimator. The green-dash trace is clearly separated by the red trace and it is providing the dominant contribution to the uncertainty of the measured spectrum. Clearly such a result is valid as soon as the values provided by the fit are accurate in the range of the fit errors.

## 6. Conclusions

The most accurate estimate of the force noise affecting LTP test masses is provided by the power spectrum of the output of the differential interferometer channel. This means that the contributions of the forces acting on the two different test masses cannot be disentangled each other. The force noise is calculated with an off-line data reduction procedure which is transforming measured displacement in effective force per unit of mass. Such an operation is based on a dynamical model of the system in which some physical parameters can be set externally. The parameters values are determined by a series of system identification experiments in which we perform least squares fits to the LTP response to known sinusoidal stimuli. The final values for the effective force noise, in the frequency range (around 1 mHz) of interest for LISA Pathfinder and LISA, are then affected by two sources of uncertainty. The first is connected with the statistical properties of the power spectral density estimator. The second arises from the propagation of the fit uncertainties on the parameters of the system model. We calculated the magnitude of the propagated fit uncertainty with a Montecarlo simulation on a set of data produced by the LISA Pathfinder simulator. We demonstrated that the dominant contribution to the force noise uncertainty comes from the statistical properties of the spectral estimator.

## References
[1]   Armano M et al. 2009 Class. Quantum Grav. 26 094001
[2]   McNamara P, Vitale S and Danzmann K 2008 Class. Quantum Grav. 25 114 034
[3]   Bortoluzzi D et al. 2004 Class. Quantum Grav. 21 S573
[4]   Bortoluzzi D et al. 2003 Class. Quantum Grav. 20 S89
[5]   LTPDA: a MATLAB toolbox for accountable and reproducible data analysis, http://www.lisa.aei-hannover.de/ltpda/.
[6]   Hewitson M et al. 2010 Class. Quantum Grav. LISA 8 special issue
[7]   Congedo G et al. 2010 JPCS. LISA 8 issue---

[8] The sample spectrum is approximately $\chi^2_2$ distributed around the expected value S. Thus, in practice, the distribution of the sample spectrum is $S\chi^2_2/2$, whereas the distribution of an average on N realization has 2N degrees of freedom and is $S\chi^2_{2N}/2N$ [10].